\documentclass[aps,prb,twocolumn,superscriptaddress]{revtex4}

\usepackage{graphicx}	
\usepackage{amssymb}	
\usepackage{amsfonts}	
\usepackage{amsmath}	
\usepackage{xcolor}		
\usepackage[T1]{fontenc}	
\usepackage[utf8]{inputenc} 
\usepackage{makecell} 
\usepackage{booktabs} 

\usepackage[english]{babel} 	

\usepackage[caption=false]{subfig}		
\usepackage{braket}		
\usepackage[load-configurations = abbreviations]{siunitx}	

\begin{document}

\title{Investigating the high-temperature thermoelectric properties of n-type rutile TiO$_2$}

\author{S. Th\'ebaud}
\email[E-mail: ]{simon.thebaud@univ-lyon1.fr}
\affiliation{Univ Lyon, Universit\'e Claude Bernard Lyon 1, CNRS}
\author{Ch. Adessi}
\affiliation{Univ Lyon, Universit\'e Claude Bernard Lyon 1, CNRS}
\author{G. Bouzerar}
\affiliation{Univ Lyon, Universit\'e Claude Bernard Lyon 1, CNRS}

\begin{abstract}
Transition metal oxides are considered promising thermoelectric materials for harvesting high-temperature waste heat due to their chemical and thermal stability, abundance and low toxicity. Despite their typically strong ionic character, they can exhibit surprisingly high power factors $\sigma S^2$, as in n-type SrTiO$_3$ for instance. Thus, it is worth examining other transition metal oxides that might equal or even surpass the performances of strontium titanate. This theoretical paper investigates the thermoelectric properties of n-type rutile TiO$_2$, which is the most stable phase of titanium oxide up to \SI{2000}{K}. The electronic structure is obtained through density functional theory calculations, while the prominent features of strong electron-phonon interaction and defects states are modelled using a small number of parameters. The thermoelectric transport properties are computed by solving the Boltzmann transport equation with the relaxation time approximation. The theoretical results are compared with a wealth of experimental data from the literature, yielding very good agreements over a wide range of carrier concentrations. This validates the hypothesis of band conduction in rutile TiO$_2$ and allows the prediction of the high-temperature thermoelectric properties.
\end{abstract}

\maketitle

\section{Introduction}

In the past decades, the global energy crisis has led to a renewed interest in the prospect of thermoelectric power generation. \cite{Zhu,Zhang2015,Zheng2011} A great deal of energy is currently wasted as heat in many industrial processes, but could be recovered and put to use if more efficient thermoelectric devices were available, particularly in manufacturing sectors involving high temperatures beyond \SI{1000}{K} such as metallurgy where furnace temperatures can be as high as \SI{2000}{K}. \cite{Forman2016,Champier2017,LeBlanc2014,Thekdi2015} When it comes to large-scale waste heat recovery, the major stumbling block remains the efficiency of thermoelectric materials. The thermoelectric performance of a given compound is measured by the figure of merit $zT$: 
\begin{equation}
zT = \frac{\sigma S^2}{\kappa} T
\end{equation}
where $\sigma$ is the electrical conductivity, $S$ is the Seebeck coefficient and $\kappa$ is the thermal conductivity. $\kappa$ is usually dominated by the phonon properties, while $\sigma S^2$, called the power factor (PF), is governed by electronic transport. The classical materials for high-temperature thermoelectrics, SiGe compounds, exhibit a $zT$ around 1. \cite{Joshi2008,Wang2008} Widespread industrial use of thermoelectric devices would require an improvement of the figure of merit by a factor 2 to 4. \cite{Bell2008,Vining2009} Since the devices are to be placed between a heat source and a heat sink, the ideal thermoelectric material should be stable and efficient over a wide range of temperature.

Recently, transition metal oxides have prompted great interest as potential thermoelectric materials. \cite{Yin2017,Ji2018,Ohta2007} They tend to be stable at high temperatures, composed of earth-abundant elements and environmentally benign. Although they are usually strongly polar, they can exhibit surprisingly high power factors. Strontium titanate, in particular, exhibits an excellent room-temperature power factor around $\SI{40}{\mu W.cm^{-1}.K^{-2}}$. \cite{Jalan2010,Bouzerar2017} However, optimization of bulk SrTiO$_3$ by doping and nanostructuring has been unable so far to yield figures of merit exceeding 0.5. Therefore, it is worth studying the thermoelectric properties of other transition metal oxides that might equal of even surpass the performances of strontium titanate. Another well-known oxide compound is TiO$_2$, which comes in three naturally occuring phases: rutile, anatase and brookite. Of these, rutile is particularly stable at high temperatures with a melting point around $\SI{2100}{K}$, while anatase and brookite undergo a phase transition to rutile near $\SI{1000}{K}$. Additionally, the carrier concentration of rutile can be tuned by doping with several elements such as boron, niobium or cobalt, or by using reduction processes to introduce oxygen vacancies acting as electron donors. \cite{Sheppard2006,Kitagawa2010,Jacimovic2013,Lu2012} Therefore, this compound would represent a very promising prospect for waste heat recovery applications if its thermoelectric properties could be optimized.\cite{Bayerl2015} 

In this paper, we investigate the electron transport properties of n-type rutile TiO$_2$ by combining first-principle calculations with a modelling of electron-phonon interactions and donor defects. This method allows us to directly compare our results with reported measurements of scattering rates, mobility, electrical conductivity, Seebeck coefficient and power factor. TiO$_2$ being a very polar material, it is known that strong interactions between electrons and optical phonons in rutile give rise to the formation of intrinsic small polarons at low temperature, i.e. the electrons are self-trapped by the surrounding deformation of the lattice. \cite{1611.06122} This behaviour has been observed both experimentally \cite{Yang2013} and in Density Functional Theory (DFT) calculations using hybrid functionals \cite{Deak2012} or DFT+U methodology. \cite{Zhao2017} On the other hand, the small polarons have been found to be highly unstable at room-temperature and above. Several DFT calculations also find these states to be only slightly favoured energetically compared to delocalized conduction states, with an energy barrier hindering the localization of extended electrons. \cite{Elmaslmane2018,Janotti2013} Furthermore, the mobility predicted by the small polaron hopping mechanism is orders of magnitudes lower than the experimental estimates in reduced samples. \cite{Deskins2007} The measured mobility above \SI{100}{K} also decreases when the temperature is elevated, \cite{Yagi1996} suggesting conduction band transport rather than small polaron hopping. Therefore, it is considered likely that the transport properties at room-temperature and above are dominated by delocalized electrons (large polarons) in the conduction band. For these reasons, we will investigate the thermoelectric properties of rutile TiO$_2$ assuming that the electrons are delocalized in the conduction band. Although the standard Generalized Gradient Approximation (GGA) functional of Perdew, Burke, and Ernzerhof \cite{Perdew1996} (PBE) is unable to describe small polaron states, it predicts the same band structure for the conduction band as the Heyd, Scuseria, and Ernzerhof \cite{Heyd2003,Krukau2006} hybrid functional, that has been used to investigate small polarons. \cite{Janotti2010} Therefore, we will compute the conduction band structure with the PBE functional, and we will introduce parameters to model the electron scattering and mass renormalization that are expected to result from strong electron-phonon coupling. These parameters will be set by direct comparison with carrier lifetime and transport measurements.

\begin{figure}
\includegraphics[width=0.9\columnwidth]{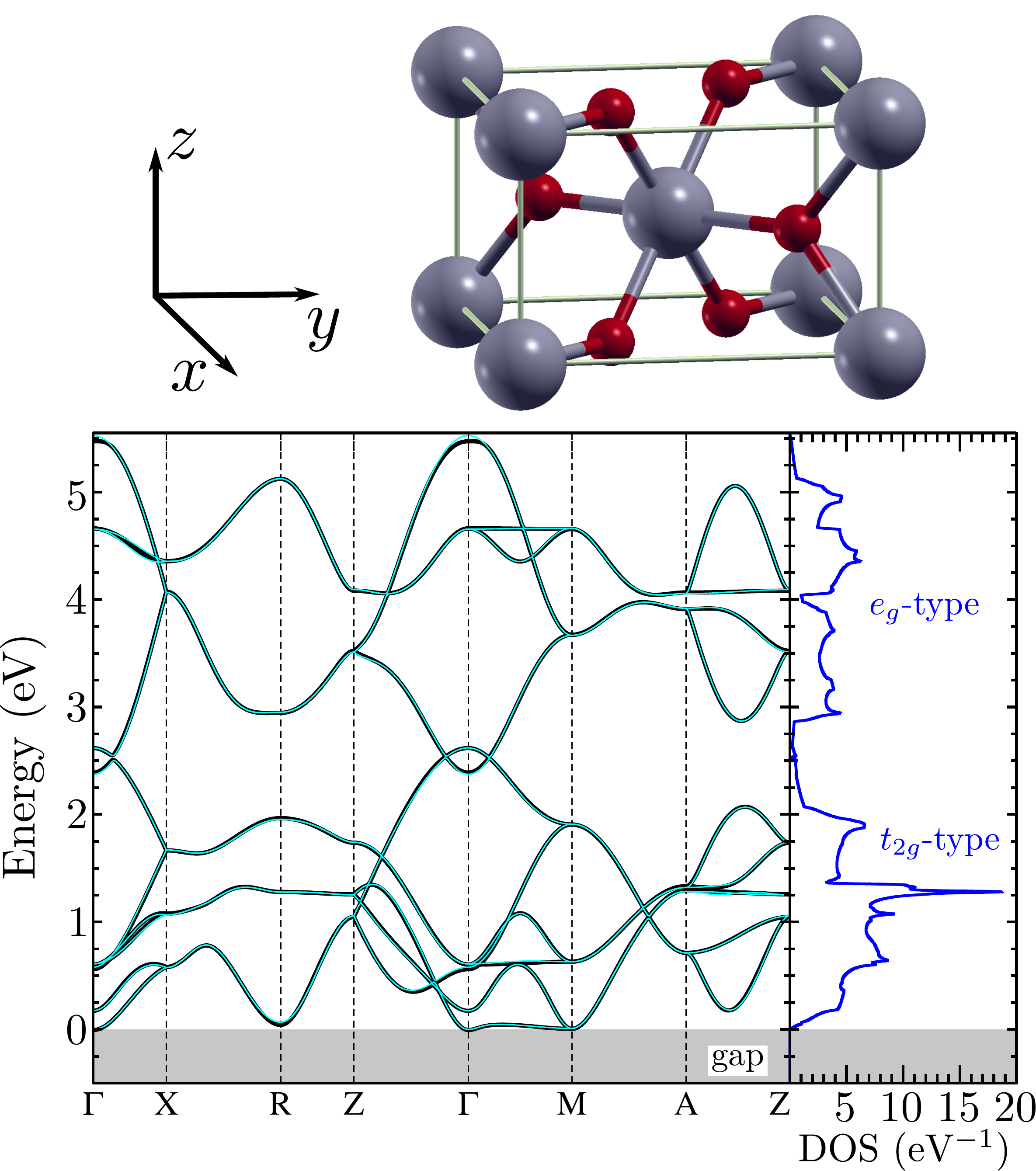}
\caption{Top: the conventional cell of rutile TiO$_2$, where the grey and red spheres represent the Ti and O atoms respectively. Bottom: The band structure and density of states of the conduction band, from full DFT calculations (black lines) and from Wannier projections (cyan lines and DOS).}
\label{fig1}
\end{figure}

We perform \textit{ab initio} calculations with the DFT package SIESTA \cite{Soler2002} on the rutile structure of TiO$_2$ (see Fig~\ref{fig1}). The GGA-PBE functional and Troullier-Martins norm-conserving pseudopotentials \cite{Troullier1991} are used. We perform an optimization of the double-$\zeta$-polarized basis with the Simplex tool of the SIESTA package. The unit cell is relaxed to forces less than $\SI{0.01}{eV/\angstrom}$ and to a pressure less than \SI{0.15}{kbar}. The self-consistent field cycles have been performed with a Monkhorst-Pack of $4 \times 4 \times 4$ k-points and a mesh cutoff of \SI{400}{Ry}. We have checked that the band-structure predicted by the plane-wave based DFT package Quantum ESPRESSO \cite{Giannozzi2009} is consistent with the results from SIESTA. The electron dispersion and the density of states (DOS) of the conduction band are shown in Fig~\ref{fig1}. The black lines corresponds to the SIESTA results, and the cyan lines and the DOS are calculated from Wannier projections of the Bloch states onto the 3d orbitals of the Ti atoms, using the Wannier90 software. \cite{Mostofi2008} The 10 Wannier orbitals can be classified as low-energy $t_{2g}$-type and high-energy $e_g$-type (more details are given in the supplemental material \cite{supplemental}), although even the latter are necessary to accurately describe the bottom of the conduction band. In the Wannier basis, the Hamiltonian matrix elements with an amplitude lower than \SI{1}{meV} have been cut, which still yields an excellent agreement with the full SIESTA band structure. Since the experimental bandgap of rutile TiO$_2$ is rather large (more than \SI{3}{eV}), a description of the valence band is unnecessary for the transport properties (we have checked that bipolar conduction is negligible in all our results). The conduction band effective mass in the $z$ direction is $m_b^z = 0.63 \, m_e$, while the effective masses in the $x$ and $y$ directions are $m_b^x = m_b^y = 1.3 \, m_e$. There is some anisotropy, but much less than in SrTiO$_3$ in which the ratio of the effective masses is as high as 10.

\section{Modeling electron transport}

We calculate the electron transport properties of n-type rutile TiO$_2$ within the framework of the Boltzmann transport equation \cite{Ashcroft,Dresselhaus2001} (BTE), using the relaxation time approximation (RTA). The electrical conductivity and Seebeck coefficients in the $i$ direction  ($i=x,y,z$) for a carrier concentration $n$ and a temperature $T$ are given by
\begin{equation}
\sigma_i(n,T) = \int dE \left( -\frac{\partial f}{\partial E}\right) \Sigma_i(E,T),
\end{equation}
and
\begin{equation}
S_i(n,T) = \frac{-1}{eT\sigma_i} \int dE \left( -\frac{\partial f}{\partial E}\right) \left( E - \mu_F \right) \Sigma_i(E,T),
\end{equation}
where $f(E,\mu_F,T)$ is the Fermi-Dirac distribution and the Fermi level $\mu_F$ is determined by integrating the DOS to find the correct electron concentration. $\Sigma_i(E,T)$ is called the transport distribution function (TDF), and its determination is key for the description of electron transport. Solving the BTE with the RTA gives
\begin{equation}
\Sigma_i(E,T) = \frac{2 e^2}{\Omega} \sum_{\mathbf{k},\lambda} v_{\mathbf{k},\lambda}^i v_{\mathbf{k},\lambda}^i \tau_{\mathbf{k},\lambda} \delta(E - \epsilon_{\mathbf{k},\lambda}),
\end{equation}
where $\Omega$ is the system size, $\mathbf{k}$ runs over the first Brillouin zone, $\lambda$ is the band index and the factor 2 accounts for spin degeneracy. $\epsilon_{\mathbf{k},\lambda}$ is the energy of the eingenstate $(\mathbf{k},\lambda)$, $v_{\mathbf{k},\lambda}^i = \frac{\partial \epsilon_{\mathbf{k},\lambda}}{\partial k_i}$ is the velocity in the $i$ direction, and $\tau_{\mathbf{k},\lambda}$ is the lifetime, considered isotropic for simplicity. The TDF is calculated using the Drude weight formalism \cite{Kohn1964,Scalapino1993} (more details are given in the supplemental material \cite{supplemental}). 

\begin{figure}
\includegraphics[width=1.0\columnwidth]{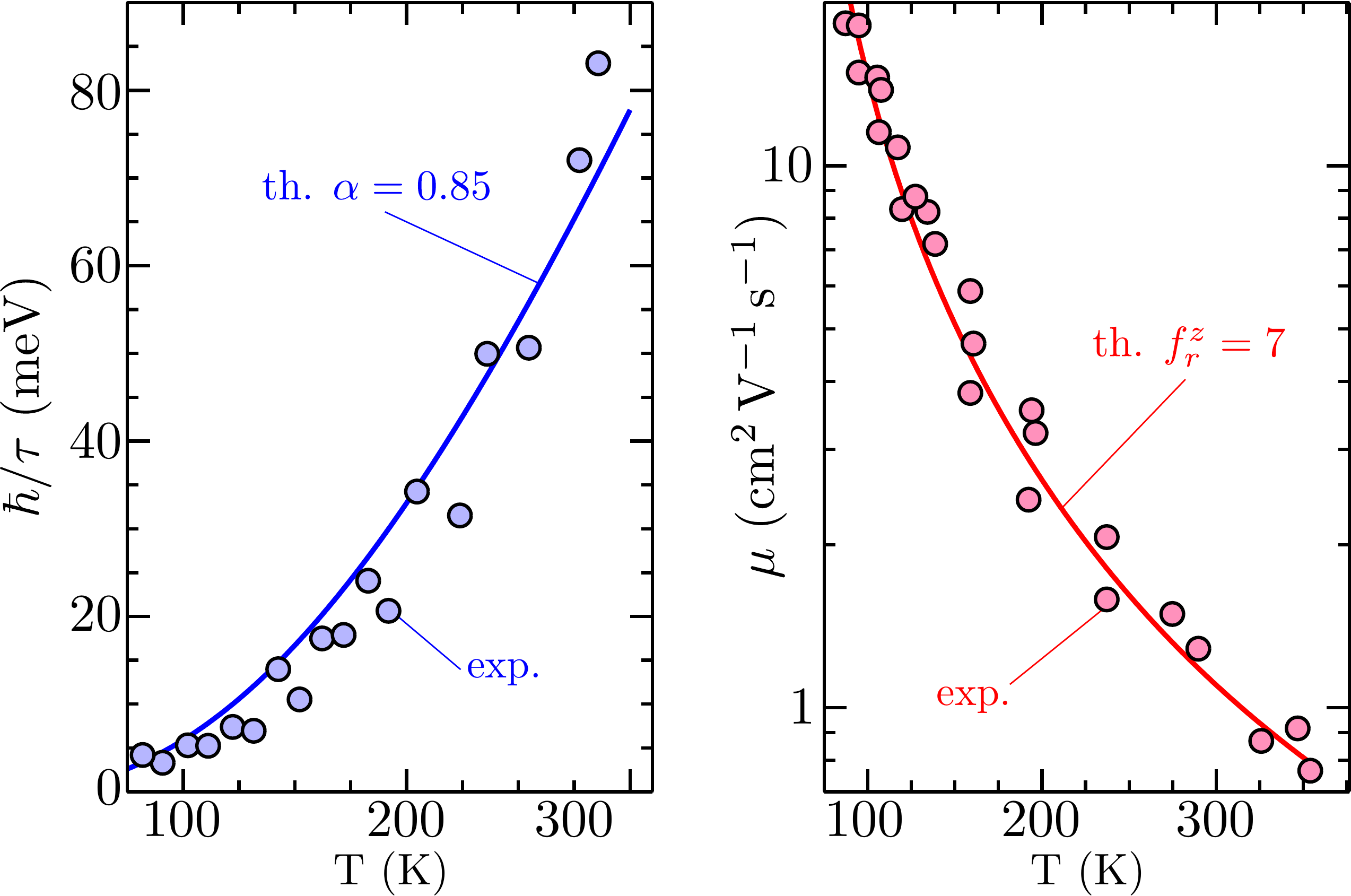}
\caption{Left: the experimental (circles, Ref.~\onlinecite{Bonn2004}) and theoretical scattering rate $\hbar/\tau$ in \si{meV}. The theoretical values correspond to the bottom of the conduction band at the $\Gamma$ point. Right: The experimental (circles, Ref.~\onlinecite{Yagi1996}) and theoretical mobility $\mu$ in the $z$ direction, in \si{cm^2.V^{1}.s^{-1}}.}
\label{fig2}
\end{figure}

The description of the thermoelectric properties requires an accurate modelling of the electron lifetime. It is especially important to get the correct temperature dependance since we are interested in high-temperature power generation. The scattering rate $\hbar/\tau$ has been measured in Ref.~\onlinecite{Bonn2004} from \SI{11}{K} to \SI{300}{K}. It increases with temperature, exhibiting rather high values around \SI{80}{meV} at room temperature (see Fig~\ref{fig2}, left panel). The temperature dependance between \SI{100}{K} and \SI{300}{K} is inconsistent with the typical $\frac{3}{2}$ power-law of the acoustic deformation potential mechanism, indeed a fit gives an exponent between $2.5$ and $3$. On the other hand, the large interaction between electrons and optical phonons can be expected to dominate scattering in such a polar compound, provided that the optical modes are sufficiently populated. \cite{Sjakste2015} First-principle calculations \cite{Lukacevic2012} and inelastic neutron scattering measurements \cite{Traylor1971} of the phonon spectrum in rutile TiO$_2$ indicate the presence of optical modes from \SI{15}{meV} to \SI{100}{meV}. Thus, longitudinal optical (LO) modes should be populated at \SI{300}{K} and above, leading to a significant polar-optical scattering rate. Assuming a parabolic dispersion for the conduction band electrons and a flat one for the LO branches, second order perturbation theory gives \cite{Mahan,Lundstrom2000}
\begin{align}
\label{scattering_law}
\frac{\hbar}{\tau_{\mathbf{k},\lambda}} & = \frac{2\alpha}{\sqrt{\epsilon_{\mathbf{k},\lambda}}} \sum_{\nu} \left( \hbar \omega_{\nu} \right)^{\frac{3}{2}} \bigg[ N_{\nu} \text{Argsh}\left( \sqrt{\frac{\epsilon_{\mathbf{k},\lambda}}{\hbar \omega_{\nu}}} \right) \\
& + \Theta \left( \epsilon_{\mathbf{k},\lambda} - \hbar \omega_{\nu} \right) \left( N_{\nu} + 1 \right) \text{Argsh}\left( \sqrt{\frac{\epsilon_{\mathbf{k},\lambda}}{\hbar \omega_{\nu}} - 1} \right) \bigg], \nonumber
\end{align}
where $\Theta$ is the Heaviside function, $\alpha$ is the Fröhlich coupling constant, $\nu$ is a branch index running over the LO branches, $\omega_{\nu}$ is the $\Gamma$-point frequency of the branch $\nu$ and $N_{\nu} = 1/(e^{\beta \hbar \omega_{\nu}}-1)$ is the Bose-Einstein occupation factor. For simplicity, we have included all optical modes in the sum and divided the resulting scattering rate by 3 (a list of the optical phonon frequencies is given in the supplemental material \cite{supplemental}). We have also assumed $\alpha$ to be the same for all modes, thus our scattering law depends only on this single parameter. To set the value of $\alpha$, we compare the calculated scattering rate at the bottom of the conduction band (see the supplemental material \cite{supplemental} for the energy dependance and full temperature dependance of $\hbar/\tau$) with the experimental values between \SI{100}{K} and \SI{300}{K}. As shown in Fig.~\ref{fig2} (left panel), the choice $\alpha = 0.85$ leads to a good agreement between theory and experiment, thus validating our assumption of dominant polar-optical electron scattering. The temperature dependance of the scattering law~\eqref{scattering_law} (see the supplemental material \cite{supplemental}) allows us to describe electron transport at high temperatures beyond \SI{300}{K}. The value $\alpha = 0.85$ corresponds to a weak Fröhlich interaction with each phonon mode, which is consistent with the use of second order perturbation theory.

Having successfully modeled the temperature-dependant scattering rate, we now endeavour to reproduce the experimental mobilities $\mu = \sigma/ne$ (with $n$ the electron density) that have been measured in Ref.~\onlinecite{Yagi1996} for the $z$~direction. However, if we calculate the TDF from the $\textit{ab initio}$ band-structure shown in Fig.~\ref{fig1}, the predicted mobility is an order of magnitude higher than the measurements, indicating a large transport mass renormalization of the delocalized carriers due to electron-phonon interactions. This could seem surprising given the relatively low coupling constant $\alpha = 0.85$, but the conduction electrons in rutile TiO$_2$ interact with five different LO branches: in such cases, the effective coupling constant $\alpha_{\text{eff}}$ for a one-branch interaction is given by $\alpha_{\text{eff}} = \sum_{\nu} \alpha_{\nu}$. \cite{Devreese2010,Verbist1992} This gives $\alpha_{\text{eff}} = 4.25$ which corresponds to the intermediate to strong coupling regime, consistent with small polarons that are slightly favoured energetically and a large mass renormalization of the delocalized conduction electrons. It is then natural to wonder why the perturbative expression for the scattering rate (equation~\eqref{scattering_law}) is validated by experiments, given the strong, non-perturbative value of the overall electron-phonon coupling constant. It must be kept in mind that, for delocalized electrons close to a metal-insulator transition, the single-particle properties (i.e. the band mass and the scattering rate obtained from the spectral function) may not be critical quantities. However, the transport mass associated with the conductivity can be much more sensitive and may be considerably enhanced due to so-called vertex corrections. \cite{Mahan}  This scenario resolves the apparent contradiction between a large mass renormalization and the delocalized behaviour exhibited by electrons in rutile TiO$_2$. To take this effect into account in the simplest way, we introduce the renormalization parameters $f_z$ and $f_x = f_y$ for transport in the $z$ and $xy$ directions, respectively. Only the TDF is renormalized: $\Sigma_i(E,T) \rightarrow \Sigma_i(E,T)/f_i$, and the renormalization parameters are adjusted so that the calculated electronic mobilities reproduce the room-temperature measurements. As shown in Fig.~\ref{fig2} (right panel), the large value $f_z = 7$ leads to an excellent agreement between theory and experiment in the whole temperature range from \SI{100}{K} to \SI{300}{K}. In the $xy$ direction, the measurements are reproduced by an even larger renormalization factor $f_x = f_y = 23.5$. These values are comparable to the effective masses found in Ref.~\onlinecite{Yagi1996} by fitting the transport measurements.  

\begin{figure}
\includegraphics[width=1.0\columnwidth]{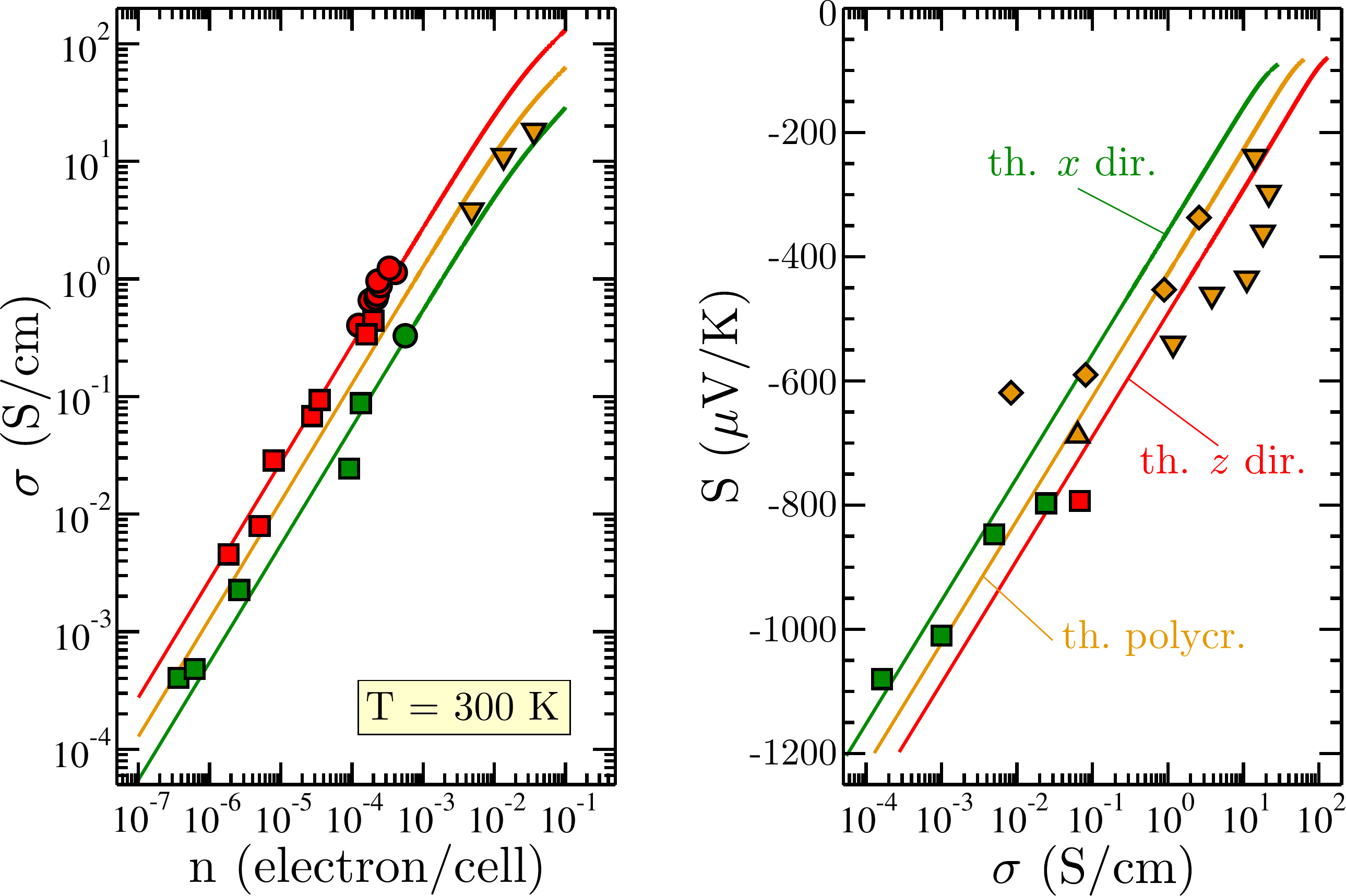}
\caption{Left: the predicted room-temperature electrical conductivity in the $z$ direction (red line), the $xy$ directions (green line) and in polycristalline samples (orange line), as a function of the carrier density $n$. Right: the predicted Seebeck coefficient as a function of the conductivity for the same directions. The experimental data are taken from Ref.~\onlinecite{Yagi1996} (circles), Ref.~\onlinecite{Acket1964,Acket1966} (squares), Ref.~\onlinecite{Kitagawa2010} (downward triangles), Ref.~\onlinecite{Jacimovic2013} (upward triangles) and Ref.~\onlinecite{Lu2012} (diamonds).}
\label{fig3}
\end{figure}

\begin{figure*}
\includegraphics[width=2.0\columnwidth]{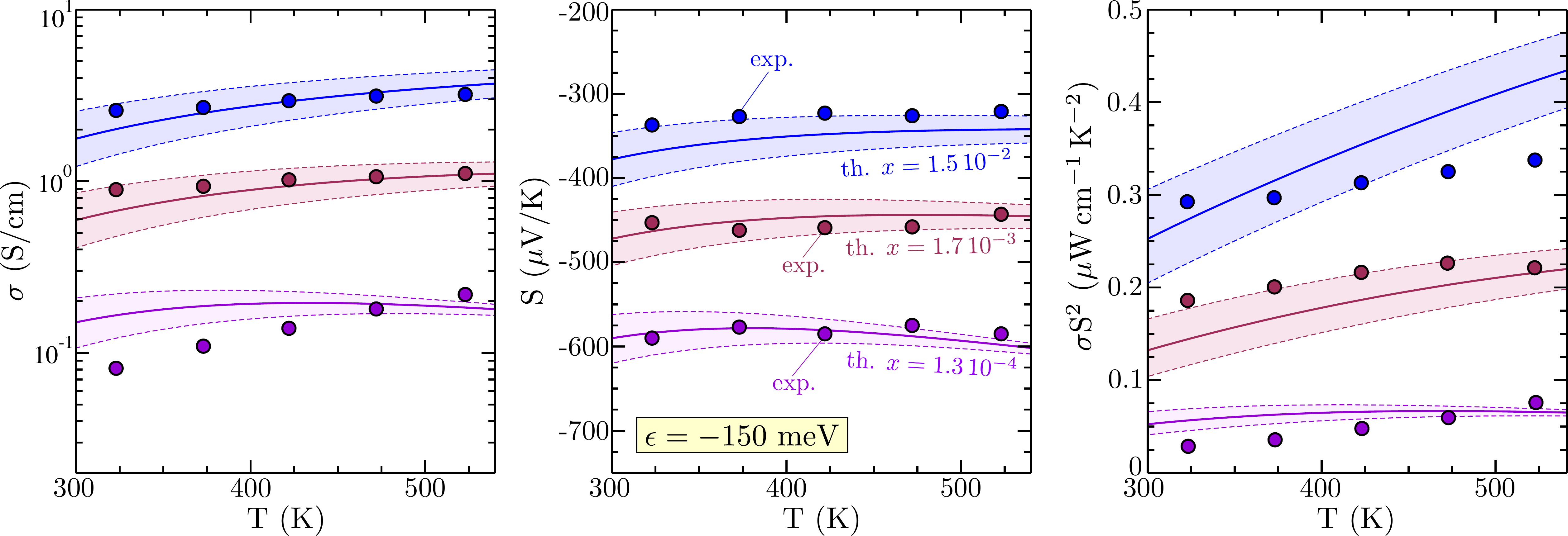}
\caption{The predicted electrical conductivity (left), Seebeck coefficient (center), and power factor (right) of polycristalline TiO$_{2-x}$ as a function of temperature for $x = \num{1.3d-4}$ (purple), $x = \num{1.7d-3}$ (red) and $x = \num{1.3d-2}$ (blue). The shaded regions correspond to a binding energy $\epsilon = -150 \pm 20$ meV. The circles are experimental data from Ref.~\onlinecite{Lu2012} for three samples subjected to SPS at \SI{1173}{K} (purple), \SI{1273}{K} (red) and \SI{1373}{K} (blue).}
\label{fig4}
\end{figure*}

The transport properties can then be calculated as functions of the electron density $n$, and compared with experimental measurements. We show in Fig.~\ref{fig3} (left panel) the room-temperature electrical conductivity in the $z$ direction (red), in the $xy$ directions (green) and in polycristalline materials (orange) for which an orientational average was taken. Experimental data from Ref.~\onlinecite{Yagi1996,Acket1964,Acket1966,Kitagawa2010} are also plotted. The agreement between theory and experiment is very good over a wide range of carrier concentration (five orders of magnitude). In the right panel of Fig.~\ref{fig3} is shown the predicted Seebeck coefficient as a function of the conductivity, together with experimental data rom Ref.~\onlinecite{Acket1964,Acket1966,Kitagawa2010,Jacimovic2013,Lu2012}. Although the theory may somewhat underestimate the Seebeck coefficient in some samples, overall it agrees well with experiments. This success of the theory at \SI{300}{K} confirms that the electronic transport properties in rutile TiO$_2$ are consistent with a band conduction mechanism based on delocalized carriers, as opposed to small polaron hopping. Moreover, it should be noted that renormalizing the DOS by the parameters $f_z$ and $f_x$ in addition to the TDF would lead to a large discrepancy between the predicted and measured Seebeck coefficients, due to the Fermi level being pushed further in the gap for a given electron concentration. This supports our hypothesis that the large mass enhancement only applies to the transport mass and that the one-particle properties can be reasonably described by perturbative expressions. 

\section{High-temperature thermoelectric performances}

As high-temperature generation is the main application prospect for TiO$_2$ as a thermolectric material, it is crucial to accurately predict the transport properties between \SI{300}{K} and \SI{2000}{K}. In Ref.~\onlinecite{Lu2012}, the electrical conductivity and Seebeck coefficients of reduced rutile samples, subjected to spark plasma sintering (SPS) at different temperatures, have been measured up to \SI{523}{K}. The conductivity displays a very weak temperature dependance, even though the scattering rate, equation~\eqref{scattering_law}, increases substantially with elevated temperature. This suggests an activation mechanism for the electron density in the conduction band, as confirmed by the Hall measurements in Ref.~\onlinecite{Yagi1996} that clearly show an increase of the carrier density with temperature. On the theoretical side, several first-principle calculations of oxygen vacancies in rutile TiO$_2$ find energetically favored electronic bound states localized on neighboring Ti atoms. \cite{Janotti2013,Zhao2017,Deak2012,Mattioli2008,Lin2015,Liu2018a} There is no consensus on the binding energy, although most experimental and theoretical estimates fall between \SI{50}{meV} and \SI{200}{meV}. The situation is similar for Ti interstitials, another important intrinsic defect, \cite{Finazzi2009,Mattioli2010} and for extrinsic dopants such as Nb and F substitutions. \cite{Janotti2013} 

To capture the effects of this activation mechanism in a simple and general way, we model the presence of oxygen vacancies by adding localized defect levels inside the gap at an energy $\epsilon$ in the DOS. Thus, TiO$_{2-x}$ is modeled by adding a term $4x\delta(E-\epsilon)$ in the pristine DOS, while the TDF is unchanged (the impurity states have zero conductivity). The total carrier density is set at $4x$ electron/cell (each oxygen vacancy is assumed to bring 2 electrons), leading to an activation mechanism with more and more carriers populating the conduction band from the defect states as temperature is increased. We adjust the parameters $\epsilon$ and $x$ to reproduce the experimental conductivities from the samples in Ref.~\onlinecite{Lu2012} subjected to SPS at \SI{1173}{K}, \SI{1273}{K} and \SI{1373}{K}. Fig.~\ref{fig4} shows the polycristalline electrical conductivity, Seebeck coefficient and power factor for $\epsilon = -\SI{150}{meV}$ and three vacancy concentrations $x = \num{1.3d-4}$, $x = \num{1.7d-3}$ and $x = \num{1.3d-2}$. These values lead to a rather good agreement with experimental data, and in particular they reproduce the weak temperature dependance of the conductivities, Seebeck coefficient and power factor, thus validating our approach. The shaded regions represent the sensibility of the results with respect to $\epsilon$, underscoring that the agreement between theory and experiment is satisfactory given the simplicity of the model (only one temperature and concentration independant impurity level). The important disparities in the predicted vacancy concentration between samples (two order of magnitudes) reflect the measured resistivities that exhibit similarly large variations. 

\begin{figure}
\includegraphics[width=0.9\columnwidth]{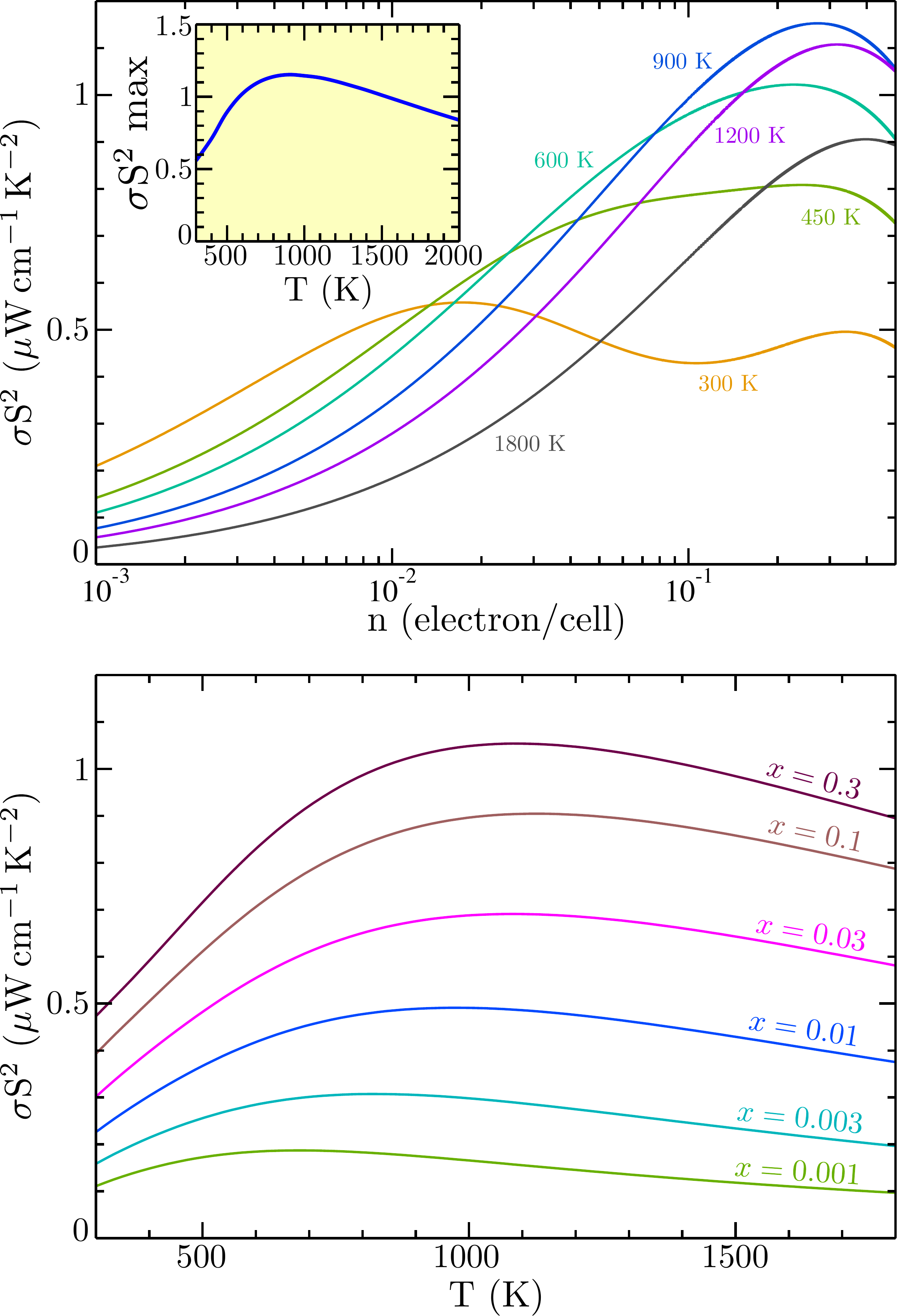}
\caption{Top: the polycristalline power factor $\sigma S^2$ as a function of the density $n$ of conduction electrons for several temperatures. Inset: the maximum PF as a function of temperature. Bottom: the PF of polycristalline TiO$_{2-x}$ as a function of temperature for several oxygen vacancy concentrations.}
\label{fig5}
\end{figure}

The thermoelectric power factor $\sigma S^2$ of rutile TiO$_2$ can now be calculated between \SI{300}{K} and \SI{2000}{K}. Fig.~\ref{fig5} shows the predicted polycristalline PF as a function of the density $n$ of conduction electrons for several temperatures (top) and as a function of temperature for several oxygen vacancy concentrations (bottom). The optimum carrier concentration is quite large around 0.2 electron/cell, corresponding to a maximum value of $\approx \SI{1.15}{\mu W.cm^{-1}.K^{-2}}$ (see the inset). Unsurprisingly, a large number of oxygen vacancies (more than $10\%$) are necessary to provide these carriers, but the activated conduction makes the power factor very stable with temperature between \SI{500}{K} and \SI{1800}{K}. Still, $\SI{1}{\mu W.cm^{-1}.K^{-2}}$ is a rather low value compared to other oxides such as SrTiO$_3$, which has an optimum power factor 40 times larger at room temperature. This poorer performance of titanium oxide is likely caused by two important features of TiO$_2$. First, and as noted previously, its band structure exhibits only modest anisotropy due to its crystal structure. In SrTiO$_3$, by contrast, the $t_{2g}$ orbitals that make up the conduction bands are oriented along the crystal axis, thus leading to a very two-dimensional character of electron transport. This is not the case in TiO$_2$. Second, the electron-phonon interactions lead to a strong mass enhancement and to a lower mobility in TiO$_2$ than in SrTiO$_3$, which is detrimental to the power factor. 

The maximum value of the figure of merit $zT_{\text{max}}$ in rutile can be roughly estimated assuming an amorphous value $\kappa \approx \SI{1}{W.m^{-1}.K^{-1}}$ for the thermal conductivity, as was done in Ref.~\onlinecite{Bayerl2015} This leads to $zT_{\text{max}} \approx 0.15$ around \SI{1800}{K}. This value is six times lower than the estimate of Ref.~\onlinecite{Bayerl2015}, which aimed at comparing the different TiO$_2$ phases and thus did not include the temperature dependance of the scattering rate. Even accounting for the occasional underestimation of the predicted Seebeck coefficient, we do not expect the figure of merit to exceed 0.6, which is insufficient for widespread applications in power generation. Therefore, if rutile TiO$_2$ is to be useful as a thermoelectric material, significant changes in its electronic structure must be engineered in order to boost the power factor beyond what can be reached by simply increasing the carrier concentration.

\section{Summary and conclusion}

In conclusion, we have investigated thermoelectric transport in n-type rutile TiO$_2$ through a combination of \textit{ab initio} calculations for the band-structure and model descriptions for the electron-phonon interaction and oxygen vacancies. The parameters for the polar-optical coupling scattering rate, mass renormalization and defect binding energy were set by a comparison between the predicted transport properties and the available experimental measurements. A very good agreement between theory and experiment is obtained over a wide range of carrier concentrations, supporting a band conduction picture of electronic transport at room-temperature and above. We predict a maximum power factor of $\SI{1.15}{\mu W.cm^{-1}.K^{-2}}$ reached at $\SI{900}{K}$ for a large carrier density of 0.2 electron/cell, which requires more than $10 \%$ oxygen vacancies in reduced samples. Such a low value of the power factor leads to an estimate of around $0.15$ for the maximum figure of merit if the thermal conductivity is reduced to its amorphous limit. Therefore, the power factor must be boosted significantly before rutile TiO$_2$ can be widely used as a thermoelectric material in power generation modules. This might possibly be achieved using quantum confinement and energy filtering effects, for instance. 

\appendix

\section*{Wannier orbitals}

\begin{figure}[h!]
\includegraphics[width=0.85\columnwidth]{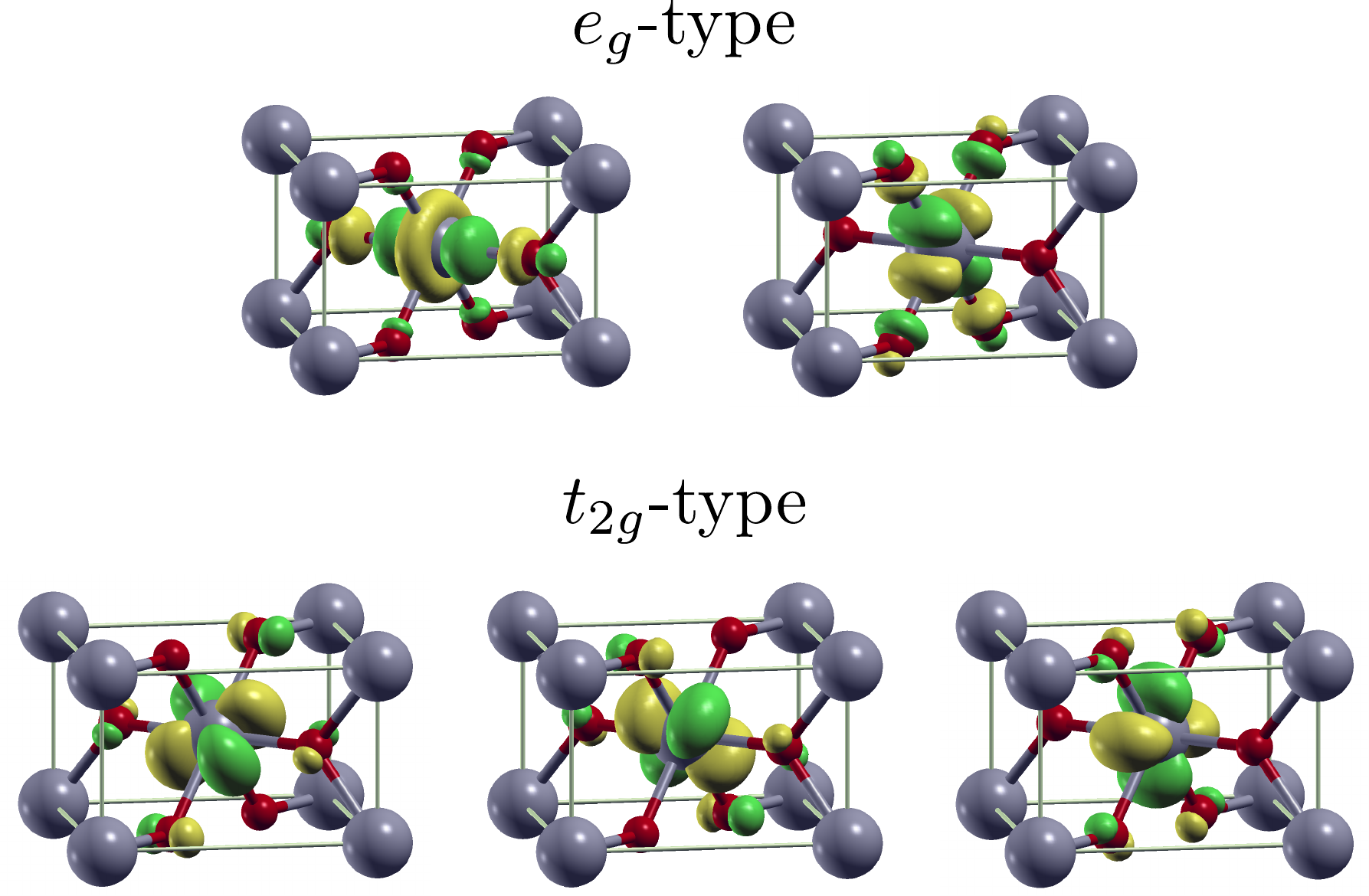}
\caption{The Wannier orbitals associated with the conduction band of rutile TiO$_2$.}
\label{fig_sm_orbitals}
\end{figure}

\begin{figure}[h!]
\includegraphics[width=0.8\columnwidth]{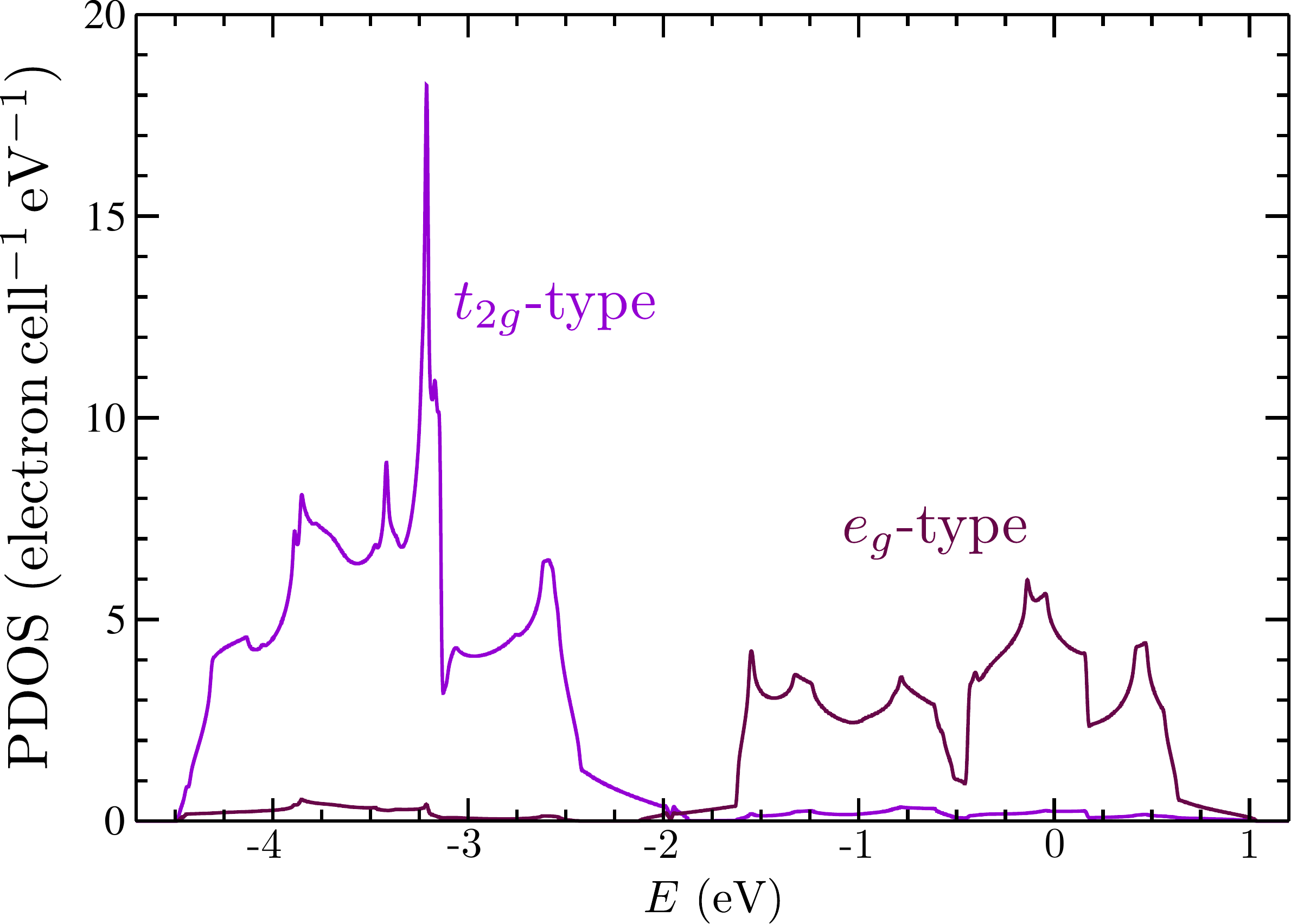}
\caption{The partial density of state (PDOS) associated with the $t_{2g}$ and $e_g$ orbitals of the conduction band.}
\label{fig_sm_pdos}
\end{figure}

In Fig.~\ref{fig_sm_orbitals} are shown the five Wannier orbitals centered on the central Ti atom of the TiO$_2$ unit cell, out of the ten orbitals that make up the conduction band of rutile TiO$_2$. They bear a strong resemblance with the 3d orbitals of the Ti atom, although there is significant weight on the 2p orbitals of the neighbouring oxygens. They can be classified as low-energy $t_{2g}$-like orbitals and high-energy $e_g$-like orbitals following the crystal field classification. The partial density of states (PDOS) of the $t_{2g}$ and $e_g$ orbitals are shown in Fig.~\ref{fig_sm_pdos}. Although the lower half of the conduction band is overwhelmingly composed of the $t_{2g}$ orbitals, there is still some weight from the $e_g$ ones at the bottom of the band, thus it is necessary to take the ten of them into account in order to properly describe the band-structure in the energy region of interest.

\section*{Drude weight}

The TDF is calculated using the Drude weight formalism \cite{Kohn1964,Scalapino1993} within the framework of the Boltzmann transport equation.\cite{Ashcroft,Dresselhaus2001} If the lifetime is a function of energy ($\tau_{\mathbf{k},\lambda} = \tau(\epsilon_{\mathbf{k},\lambda})$), the TDF in the $i$ direction can be written $\Sigma_i(E) = D_{0,i}(E) \tau(E)$ with
\begin{equation}
D_{i}(E) = \frac{2 e^2}{\Omega} \sum_{\mathbf{k},\lambda} v_{\mathbf{k},\lambda}^i v_{\mathbf{k},\lambda}^i \delta(E - \epsilon_{\mathbf{k},\lambda}).
\end{equation}
$D_{i}(E)$ is called the Drude weight. It can be re-expressed as
\begin{equation}
\label{drude_flux}
D_{i}(E) = \frac{2 e^2}{\hbar^2 \Omega} \sum_{\epsilon_{\mathbf{k},\lambda} < E} \frac{\partial^2 \epsilon_{\mathbf{k},\lambda}}{\partial k_i^2} = \frac{2}{\Omega} \sum_{\epsilon_{\mathbf{k},\lambda} < E} \frac{\partial^2 \epsilon_{\mathbf{k},\lambda}}{\partial A_i^2}, 
\end{equation}
where $A_i$ is a uniform vector potential in the $i$ direction introduced through the Peierls substitution. We use the last expression in equation~\eqref{drude_flux} to calculate the Drude weight in practice. 

\begin{figure}[h]
\includegraphics[width=0.8\columnwidth]{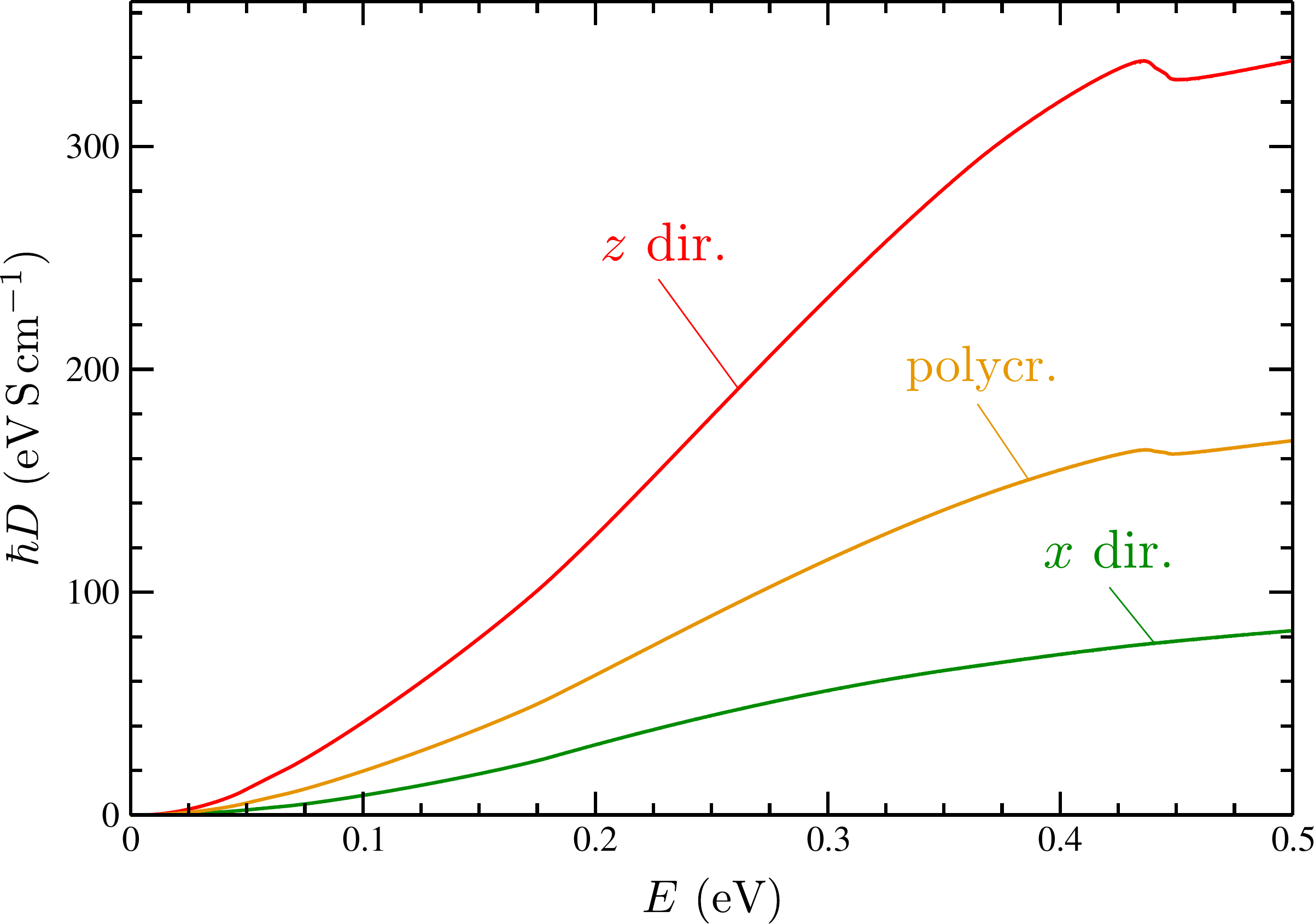}
\caption{Renormalized Drude weights as a function of energy in the $xy$ directions, $z$ direction and for polycristalline samples.}
\label{fig_sm_drude_weight}
\end{figure}

In the parabolic approximation, the Drude weight reduces to $D_{i}(E) = \frac{n e^2}{m_i}$, with $n$ the carrier density. Thus, to take into account the mass enhancement due to electron-phonon interaction, we renormalize the Drude weight: $D_i(E) \rightarrow D_i(E)/f_i$. The renormalized Drude weights in the $xy$ directions, in the $z$ direction and for polycristalline samples ($D_{\text{polycr}} = (2 D_x + D_z)/3$) are shown in Fig.~\ref{fig_sm_drude_weight}.

\vspace{2cm}

\section*{Electron-phonon scattering rate}

\begin{figure}[h!]
    \centering
        \subfloat[The scattering rate as a function of energy for several temperatures.]{\includegraphics[width=0.8\columnwidth]{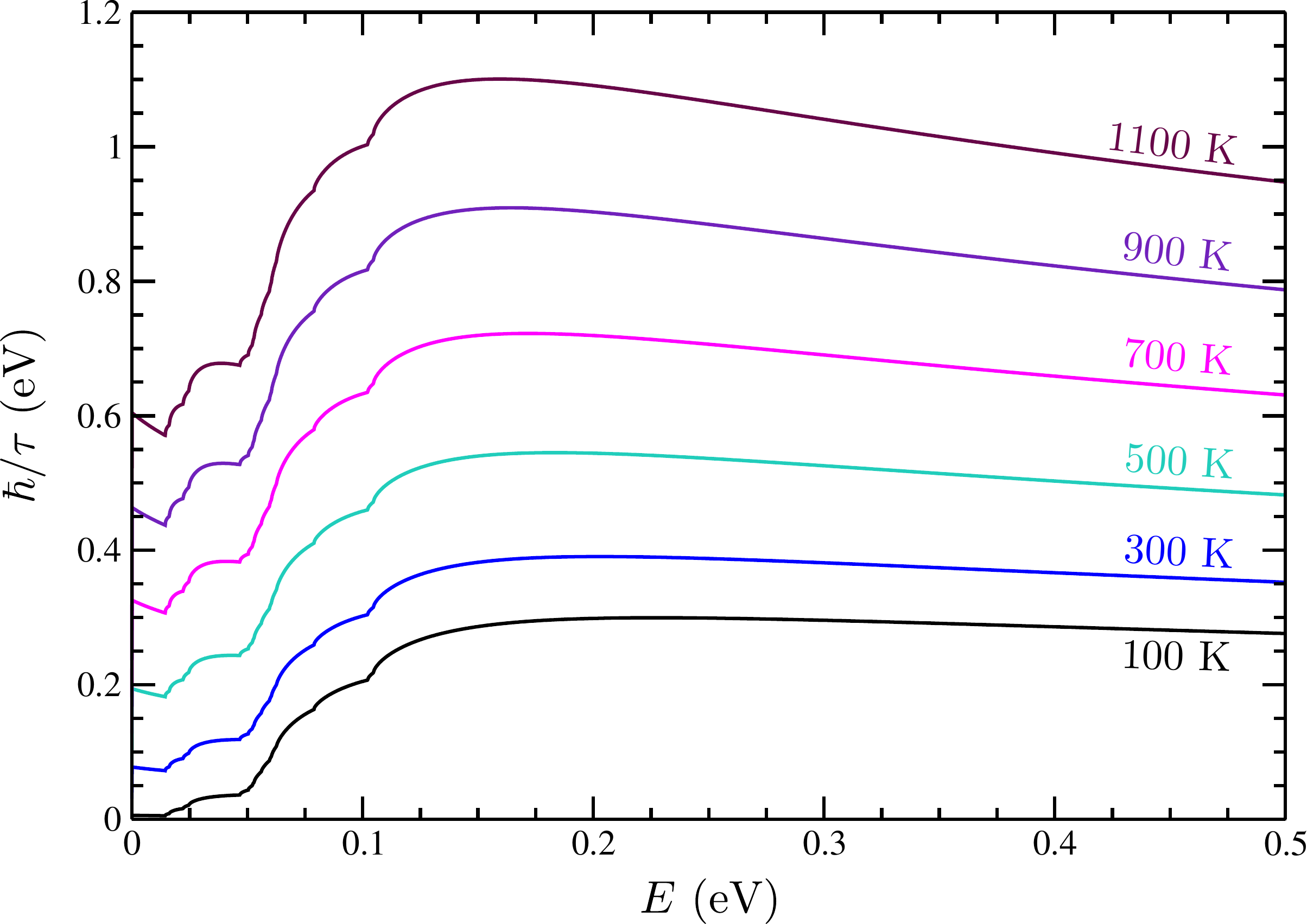} \label{fig_sm_scattering_E}}
        \qquad
        \subfloat[The scattering rate at the $\Gamma$ point as a function of temperature.]{\includegraphics[width=0.8\columnwidth]{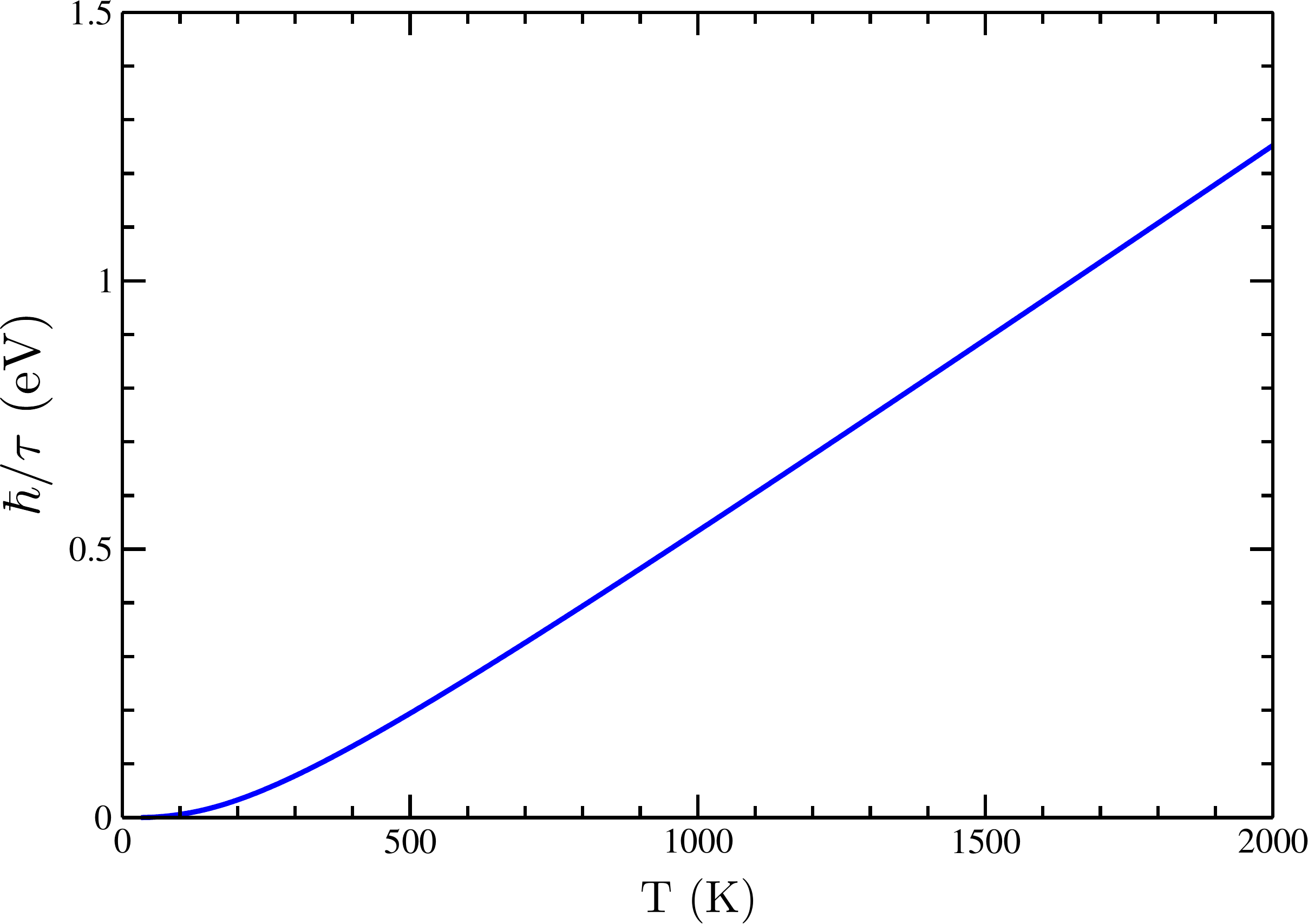} \label{fig_sm_scattering_T}}
    \caption{The electron-phonon scattering rate.}
\end{figure}

The electron scattering rate due to collisions with optical phonons is calculated for the Fröhlich interaction Hamiltonian, i. e. assuming a parabolic dispersion for the conduction band electrons and a flat dispersion for the LO branches. The first diagram for the electron self-energy leads to the scattering rate \cite{Mahan,Lundstrom2000}
\begin{align}
\label{scattering_law_sm}
\frac{\hbar}{\tau_{\mathbf{k},\lambda}} & = \frac{2\alpha}{\sqrt{\epsilon_{\mathbf{k},\lambda}}} \sum_{\nu} \left( \hbar \omega_{\nu} \right)^{\frac{3}{2}} \bigg[ N_{\nu} \text{Argsh}\left( \sqrt{\frac{\epsilon_{\mathbf{k},\lambda}}{\hbar \omega_{\nu}}} \right) \\
& + \Theta \left( \epsilon_{\mathbf{k},\lambda} - \hbar \omega_{\nu} \right) \left( N_{\nu} + 1 \right) \text{Argsh}\left( \sqrt{\frac{\epsilon_{\mathbf{k},\lambda}}{\hbar \omega_{\nu}} - 1} \right) \bigg], \nonumber
\end{align}
where $\Theta$ is the Heaviside function, $\alpha$ is the Fröhlich coupling constant, $\nu$ is a branch index running over the LO branches, $\omega_{\nu}$ is the $\Gamma$-point frequency of the branch $\nu$ and $N_{\nu} = 1/(e^{\beta \hbar \omega_{\nu}}-1)$ is the Bose-Einstein occupation factor. The first term inside the bracket corresponds to the absorption of an optical phonon, while the second term is associated with the emission of an optical phonon and thus requires the electron energy to be higher than the optical mode frequency. For simplicity, we have included all optical modes in the sum of equation~\eqref{scattering_law_sm} and divided the result by 3. The optical mode frequencies, extracted from Ref.~\onlinecite{Lukacevic2012}, are listed in Table~\ref{table_modes_optiques}. The scattering rate as a function of energy is shown in Fig.~\ref{fig_sm_scattering_E} for several temperatures. The discontinuities in the derivative are caused by the absorption processes that activate when the energy reaches the phonon mode frequency. The scattering rate at the $\Gamma$ point is shown in Fig.~\ref{fig_sm_scattering_T} as a function of temperature up to \SI{2000}{K}. It becomes linear above \SI{500}{K} as the optical modes are more and more populated.

\begin{table}[h]
\vspace{5pt}
\setlength{\tabcolsep}{10pt}
\resizebox{0.45\columnwidth}{!}{
\begin{tabular}{ c c } 
branch & $\hbar \omega$ (meV) \\
\toprule
1 & 14.4  \\
\midrule
2 & 16.2 \\
\midrule
3 & 22.2  \\
\midrule
4 & 24.7  \\
\midrule
5 & 46.8 \\
\midrule
6 & 50.4  \\
\midrule
7 & 52.2  \\
\midrule
8 & 53.0 \\
\midrule
9 & 56.1  \\
\midrule
10 & 59.6  \\
\midrule
11 & 60.5 \\
\midrule
12 & 62.7  \\
\midrule
13 & 78.9  \\
\midrule
14 & 102.2  \\
\midrule
15 & 104.7  \\
\bottomrule
\end{tabular}
}
\caption{The optical mode frequencies used in the calculation of the electron scattering rate.}
\label{table_modes_optiques}
\end{table}

\begin{footnotesize}

\end{footnotesize}

\end{document}